\documentclass[iop]{emulateapj}

\slugcomment{Invited Spotlight Article for IRSN Astronomy and Astrophysics}
\shortauthors{Zaritsky}
\shorttitle{Scaling Relations}

\begin{document}
\title{Implications and Applications of Kinematic Galaxy Scaling Relations}
  
\author{Dennis Zaritsky}
\affil{Steward Observatory, University of Arizona, 933 North Cherry Avenue, Tucson, AZ 85721}
\email{dennis.zaritsky@gmail.com}

\begin{abstract}   
Galaxy scaling relations, which describe a connection between ostensibly unrelated physical characteristics of galaxies, testify to an underlying order in galaxy formation that requires understanding. I review the development of a scaling relation that 1) unites the well-known Fundamental Plane (FP) relation of giant elliptical galaxies and Tully-Fisher (TF) relation of disk galaxies, 2) fits low mass spheroidal galaxies, including the ultra-faint satellites of our Galaxy, 3) explains the apparent shift of lenticular (S0) galaxies relative to both FP or TF, 3) describes all stellar dynamical systems, including systems with no dark matter (stellar clusters), 4)  associates explicitly the numerical coefficients that account for the apparent ``tilt" of the FP away from the direct expectation drawn from the virial theorem with systematic variations in the total mass-to-light ratio of galaxies within the half-light radius, 5) connects with independent results that demonstrate the robustness of mass estimators when applied at the half-light radius, and 6) results in smaller scatter for disk galaxies than the TF relation.  The relation develops naturally from the virial theorem, but implies the existence of  additional galaxy formation physics that must now be a focus of galaxy formation studies. More pragmatically, the relation provides a lynchpin that can be used to measure distances and galaxy masses. I review two applications: 1) the cross-calibration of distance measurement methods, and 2) the determination of mass-to-light ratios of simple stellar populations as a function of age, and implications of the latter for the stellar initial mass function.
\end{abstract}

\section{Introduction}

To develop an understanding of any set of objects, we first  classify them in the expectation that this will help us uncover the rules that describe the set. For stars, this systematic approach led to stellar classification, eventually to the Hertzprung-Russell diagram, and finally to theories of stellar structure and nuclear burning that comprise one of astronomy's fundamental successes of the previous century. For galaxies, this approach has been less successful  in uncovering simple intuitive guiding principles. In part, this failure was due to the absence of a comprehensive description of galaxy structure akin to that available for stars. Theories of galaxy formation, currently represented mostly by numerical simulations \citep[for example see][]{springel}, are left to describe a loosely tied set of observables that include galaxy luminosity functions, clustering properties, color distributions, and star formation rates for ensembles of systems, rather than  the specific characteristics of any individual galaxy. As such, even if these models successfully reproduce the existing ensemble observations, our understanding of galaxies would be quite different in nature than our understanding of stars.

What does this long-running failure to identify simple rules of galactic structure signify?
Perhaps it reflects a greater underlying complexity to galaxies than to stars. Perhaps the formation and evolution of galaxies is so strongly sensitive to different variables that each galaxy is an entirely distinct entity and we will never find a simple description of galaxy structure that is both broadly applicable and sufficiently precise for individual galaxies. 
Seen in this light, it becomes clear that the search for the unifying principles of galactic structure is in essence an attempt to determine the degree to which simple, intuitive guiding principles for galaxy formation can accurately describe real galaxies at a non-trivial level of detail.
At its core, this is an argument between the potential value of an analytic description of galaxy formation vs. the need for numerical simulations. 

In this article, I describe recent work that has demonstrated that there is indeed an underlying simple order to stellar systems of all types and masses beyond that which we can currently explain. The observed low scatter about this empirical relationship, which ties together the basic measurable properties of galaxies, attests to the presence of underlying rules. Perhaps the low scatter arises from a galaxy formation version of the central limit theorem or, perhaps, it points to a more intuitively meaningful connection between the way stars form and are packed within dark matter potentials. I frame the discussion of the relationship in terms of assumptions and refinements to the virial theorem as applied to galaxies to clarify the additional constraints and information provided by the empirical findings. I will show how this new scaling relation helps address a number of open questions, including some regarding the nature of S0 and low luminosity galaxies, while at the same time being applicable to all of the galaxies that were well described by the Fundamental Plane (FP) and Tully-Fisher (TF) relations. 
After describing this scaling relationship, 
which is referred to as the Fundamental Manifold (FM), in reference to its antecedent the FP,
I proceed to describe ways in which  it can be exploited to measure other quantities of core astrophysical interest, distances and galaxy masses.

\vfill\eject
\section{A Logical Path to Galaxy Scaling Relations}
\label{sec:intro}

Over time, a disjoint set of rules regarding galaxy structure, generally referred to as
scaling relations, have been identified \citep[for some examples see][]{fj,kormendy, burstein, kh, graham}. The two such relations that are in most common use, and include measurements of the internal kinematics of galaxies, are the 
Tully-Fisher relationship \citep[TF;][]{tf}  and the Fundamental Plane \citep[FP;][]{dd87,d87}. 
Quantitatively, however, these are somewhat arcane parameterizations,  with
non-integer coefficients that are derived empirically and depend on observational details such as filter passbands \citep[for example see][]{pizagno}. Qualitatively, these differ from one another in that they apply to restricted, non-overlapping sets of galaxy types and are functions of different measured quantities. These relations are therefore at best partial answers in our quest for a comprehensive description of galactic structure. As usual, with the benefit of hindsight, one can rework a clearer narrative.
We now trace a straightforward, contiguous path to the FP and TF that will explain certain features of those relationships, and to the unification of those relationships into one proposed as being applicable to all stellar systems \citep{zgz,zzg}.

\subsection{Starting From the Virial Theorem}

It is a common misconception that the scaling relations are simply rephrased versions of the virial theorem. Although their origin lies with that theorem, their existence implies additional, non-trivial, physical constraints on the nature of galaxy formation. This assertion is clarified by expressing the virial theorem in a form reminiscent of the FP: 
\begin{equation}
\log r_0 = 2 \log V_0 - \log I_0 - \log \Upsilon_0 + \log A_0 - \log B_0 - C_0,
\label{eq:fm1}
\end{equation}
where the subscript 0 indicates quantities measured at a selected radius, $r_0$: $V_0$ is a measure of the internal motions within that radius (typically either the circular velocity, velocity dispersion, or some combination), $I_0$ is the surface brightness within $r_0$, $\Upsilon_0$ is the mass-to-light ratio of the matter within $r_0$, $A_0$ and $B_0$ are coefficients arising from the integration of the kinetic and potential energy terms in the virial theorem (setting, for example, $ \int_0^\infty{v^2 r^2dr} \equiv AV^2$), and finally $C_0$ is an integration constant.
In principle, $r_0$ should be selected to encompass the entire system because the virial theorem applies to the system as a whole. However, in practice, because galaxies have no well-defined edges and the low surface brightnesses of their outskirts make measuring these quantities difficult, $r_0$ is selected to be a compromise radius, such as that which encompasses half the light of the system, $r_h$. This is the first step away from the formal (physically correct) application of the virial theorem. 

Additional assumptions and simplifications are necessary to apply Eq. \ref{eq:fm1} to real systems.
Because $A$, $B$, $C$, and $\Upsilon$ can vary from system to system, and also for different enclosed radii within the same system, there is no {\sl a priori} assurance that Equation \ref{eq:fm1} defines a simple, limited distribution of galaxies within the ($r_0$,$I_0$,$V_0$) space, aka a {\sl scaling relation}. In other words, solutions of Equation \ref{eq:fm1} exist for any combination of ($r_0$,$I_0$,$V_0$) if $\Upsilon_0$, $A_0$, $B_0$, and $C_0$ are unconstrained, and yet galaxies do not populate the entire $(r_0,I_0,V_0)$-space. The more confined the distribution of galaxies within this space, the more restrictive the constraints on the models. The value of a scaling relation is that it quantifies the degree to which nature is limiting combinations of these parameters, and, by implication, underlying additional physics that we have yet to appreciate that lies beyond the virial theorem. If galaxies are found only in limited combinations of 
$r_0$, $I_0$, and $V_0$, then it must also be true that only certain combinations of $A_0$, $B_0$, $C_0$ and $\Upsilon_0$ are allowed. Why?
 
\subsection{A Key Simplification}

I now take a slight detour in our quest for a comprehensive scaling relation by considering an important result regarding the measurement of galaxy masses. 
The virial theorem, and hence Equation \ref{eq:fm1} in a slightly different guise, is also the primary pathway to galaxy mass determinations because it can be used to measure $\Upsilon$. The difficulty that underlies all such discussions \citep[for examples see,][]{page, bahcall, heisler, erickson, little, zw} is evident from our expression of the virial theorem in Equation \ref{eq:fm1}, namely the unknown numerical values of $A$, $B$, and $C$, hereafter referred to as the virial coefficients. For simple geometries these coefficients can be evaluated analytically, for example the gravitational potential energy for a uniform density sphere is $3GM^2/5R$, leading to $B = 3G/5$ in this particular example. 
However, in reality these coefficients
are particularly troublesome because we have no way to calculate them without having a full knowledge of the gravitational potential and the tracer particle distribution function and no guarantee that whatever values we adopt remain at least roughly constant from system to system.

In the face of such ignorance, one usually adopts the simplest possibility --- that these values are the same from galaxy to galaxy --- and then proceeds to use simple modeling or back-of-the-envelope arguments to obtain numerical values that are inserted into the analogs of Eq. \ref{eq:fm1} \citep[often referred to as mass estimators:][]{bahcall, zw}.  In certain cases, additional data, such as measurements of the higher order moments of the line of sight velocity distribution, constrain the orbits of the tracer particles \citep{smith,dejonghe,merrifield}, providing independent information on the virial coefficiencts. These approaches typically center on the application of the Jeans equation \citep[for example see][]{merrifield} or Schwarschild modeling \citep[for a few examples see][]{vdv,statler,cretton} to help in the interpretation of the data, but both require data that is far superior than what is typically available. Nevertheless, such studies provide critical tests for any less sophisticated method, assuming that such methods are not inherently doomed by system to system variations in the values of the virial coefficients. The key prerequisite to the use of mass estimators is establishing that variations in 
the virial coefficients are not the dominant source of uncertainty.

\cite{walker} and \cite{wolf} demonstrated, using a range of dynamical models for spheroidal galaxies, that the enclosed mass {\it at the half light radius}, $M_h$, as estimated from easily measured observables (size, luminosity, and velocity dispersion), is robust to the unknown details of the internal kinematics and structure of the stellar system \citep[robust to $\sim 10$\% or 0.05 dex,][]{wolf}.
Those studies were motivated by the desire to measure accurate and precise masses for low mass stellar systems as tests of hierarchical structure formation models and dark matter halo profiles\footnote{\cite{wolf} argue further that the true half-mass radius, rather than the projected half-mass radius works best, while \cite{walker} present an analysis in projected space. The two results are consistent and  we opt to use the \cite{walker} result which involves projecting models rather than deprojecting the observations.}. However, in the language of Equation \ref{eq:fm1}, those studies show that the virial coefficients are materially identical from system to system --- {\sl if one applies the equation at the half-light radius}, $r_h$. Correspondingly, they also show that if one attempts to use analogs of Eq. \ref{eq:fm1} for quantities measured at radii other than $r_h$ one will realize more scatter in the mass estimates. This work demonstrates that the success of the FP, and the extended scaling relation we discuss here, lies in large part to a fortuitous choice of $r_h$ as the scaling relation's fiducial radius. The value of the \cite{walker} and \cite{wolf} work, in the context of the current discussion, is that it codified what had generally been assumed without much supporting evidence and also highlighted when and how the assumption breaks down.

Using the results of that work, I calculate a numerical value for the combined quantity $\log A_h - \log B_h -  C_h$ in Equation \ref{eq:fm1}.
\cite{walker} find that $M_{h} = 580 r_h \sigma_v^2$, where the mass is in solar masses, $r_h$ is in pc, and $\sigma_v$ is the line of sight velocity dispersion\footnote{We explicitly add the subscript $v$ to $\sigma$ to specify that it represents a {\sl velocity} dispersion. However, in cases where we want to highlight some other aspect of $\sigma$, for example that it is measured within the half-light radius, $r_h$, we will drop the subscript $v$ and replace it with the subscript $h$, as in $\sigma_h$. Nevertheless,  $\sigma$ always refers to the line-of-sight velocity dispersion if it carries any subscript.}  in km s$^{-1}$. Rewriting this expression in a form similar to Eq. 1 by defining $I_h \equiv L_{h}/\pi r_h^2$ and converting units, 
results in
\begin{equation}
\log r_h = 2 \log \sigma_v - \log I_h - \log \Upsilon_{h} - 0.73
\label{eq:walker}
\end{equation}
comparing Eqns. \ref{eq:fm1} and \ref{eq:walker} highlights the obvious similarity and leads us to  associate $\sigma_v$ with $V$ for spheroidal galaxies and conclude that $\log A_h - \log B_h - C_h = - 0.73$. The derived value of the combined virial coefficients in Equation \ref{eq:walker} can be tested by comparing mass estimates within $r_h$ obtained using the more robust methods (Jeans or Schwarschild modeling and/or gravitational lensing model \citep{slacs}) to those obtained by applying Equation \ref{eq:walker} to get $\Upsilon_h$ and then multiplying by the luminosity $L_h$. Following that approach, \cite{zzg} independently (prior to the \cite{walker} study) found a combined value of the virial coefficients of $-0.75$, in what turned out to be excellent agreement with Equation \ref{eq:walker}. Confirmation that the combined virial coefficients are roughly the same from system to system comes form the low scatter about Equation 2 of real systems with independently determined values of $\Upsilon_h$ \citep[see][]{zzg}.  

Once the system-to-system stability of the virial coefficients is confirmed and accepted, the last remaining unknown in  Eq. \ref{eq:walker} is $\Upsilon_h$. Observationally, the scatter about the Fundamental Manifold (FM) is limited to $\sim$ 0.1 dex, and is even smaller for subsamples of galaxies drawn from the individual studies that comprise the heterogenous dataset in that work \citep{zzg}. However, even if the scatter were zero, it would still be the case mathematically,  that any combination of $(r_h, \sigma_h, I_h)$ would be allowed by Equation \ref{eq:walker} if $\Upsilon_h$ is unconstrained\footnote{The only obvious physical constraint on $\Upsilon_h$ is the lower limit defined by the mass-to-light ratio of a purely stellar population.}. The existence of a scaling relation, where galaxies populate a very limited region of $(r_h,\sigma_h, I_h)$-space also implies that $\Upsilon_h$ is constrained.

For dark matter free systems, the structure is now entirely defined if one evaluates $\Upsilon_h$ using simple stellar population models. Because most of the stellar clusters for which the set of $(r_h, \sigma_h, I_h)$ exist are old ($>$ 10 Gyr),  they should have nearly the same value of $\Upsilon_h$ (even if they are all of the same age, variations in $\Upsilon_h$ will exist due to chemical abundance variations and dynamical evolution). With $\Upsilon_h$ set to a constant, Equation \ref{eq:walker} describes a plane in the $\log r_h-\log \sigma_h-\log I_h$-space. Indeed, Milky Way globular clusters not only fall onto a plane, they fall onto a line \citep{pas,zzgb}, which suggests even further constraints on their structure, specifically an underlying relationship between two of the three measured parameters that removes an additional degree of freedom\footnote{Qualitatively such constraints are not difficult to imagine. For example, a cluster with extremely low surface density might not have been able to form simply because the cloud from which it would have formed from would have been tidally disrupted. Nevertheless, the quantitative constraint on formation models provided by Eq. \ref{eq:walker} could be quite challenging and informative}.

\subsection{Onward to the Fundamental Plane}

Giant elliptical galaxies empirically obey the FP, which has the form
\begin{equation}
\log r_h = \beta \log \sigma_v - \gamma \log I_h + \delta,
\label{eq:fp}
\end{equation}
where $\beta$, $\gamma$, and $\delta$ are numerical coefficients. This relationship lacks the troublesome $\Upsilon_h$ that is included in Eq. \ref{eq:walker}.
Within the framework of Eq. \ref{eq:walker}, the validity of the FP therefore requires that $\Upsilon_h \propto \sigma_v^{2-\beta}I_h^{1+\gamma}$, for those galaxies that occupy the FP.
One particular published fit to the FP, $r_h \propto \sigma_v^{1.2\pm 0.07}I_h^{-0.82 \pm 0.02}$ from \cite{cappellari}, therefore implies that $2-\beta = 0.8 \pm 0.07$ and $1+\gamma = -0.18 \pm 0.02$, or expressed in another form, that $\Upsilon_h \propto \sigma_v^{0.8} I_h^{-0.18}$ for the giant elliptical galaxies that satisfy the FP. Although alternative fits of the FP exist using different samples of ellipticals that are often observed in different filter pass bands \citep[for examples see][]{jorg,bernardi03}, they differ in detail rather than in spirit. All of these relations implicitly require that the mass-to-light ratio have power-law dependence on $\sigma_v$ and $I_h$. 

\begin{figure}
\plotone{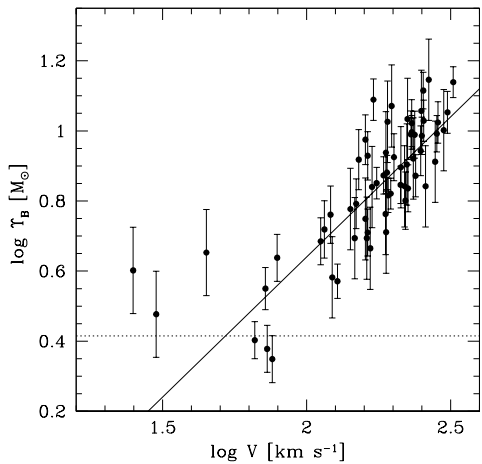}
\caption{$\Upsilon$ in the B-band for local galaxies from \cite{vdm2} using a Jeans equation analysis. Here we take $V = \sigma_v$
and have plotted all 62 of their
systems. The dotted line represents an estimate of $\Upsilon$ for a purely stellar population. The solid line represents the expectation based on the $\sigma_v$ dependence inferred from the \cite{cappellari} FP.}
\label{fig:m2l}
\end{figure}

The relationship between the total mass-to-light ratio and the structural properties of galaxies must arise from physics that dictates
how luminous baryons settle into dark matter potential wells. As such, it is critical to understand if this relationship holds for more than giant elliptical galaxies. 
Using direct measurements of $\Upsilon_h$ from Jeans modeling \citep[Figure \ref{fig:m2l}; data from][]{vdm2} we
see both that the power law relationship between $\Upsilon_h$ and $\sigma_v$ holds for giant ellipticals consistent with the slope inferred above, 0.8, as it must due to their obeying the FP,  and that it breaks down for low $\sigma_v$ ellipticals. It is not surprising that the power-law description breaks down for low $\sigma_v$ because extrapolating the power law leads to unphysical values of $\Upsilon_h$ that are smaller than those of a dark-matter-free stellar population (the dotted line in the Figure).
On the basis of these data, the FP is manifestly only valid for galaxies above a threshold $\sigma_v \sim 100$ km sec$^{-1}$. The full relationship between $\Upsilon_h$ and ($\sigma_v$, $I_h$) must therefore be more complex
than a power-law. In addition to the flattening at low $\sigma_v$ seen in Figure \ref{fig:m2l}, there is also
evidence for a turnover, or deviations from the FP, at large values of $\sigma_v$ \citep{zzg,bernardi11}.

The FP is not an all-inclusive scaling relation, even considering only spheroidal galaxies. 
The relationship between $\Upsilon_h$ and $\sigma_v$ must be of higher complexity that presumed in the FP \citep{zzg,tollerud}. The FP applies only where a power law description is an acceptable approximation to this more complicated relationship. Although the simple power-law description of  $\Upsilon_h$ fails, this failure is not a conceptual problem because there was never any physical motivation for such straightforward behavior. However, $\Upsilon_h$ can be described as a function of $\sigma_v$ even when $\sigma_v$ is $<$ 100 km sec$^{-1}$ if one is will to consider a more complex analytic description for the structure of galaxies.

\subsection{Incrementally Adding Rotation}

\begin{figure}
\plotone{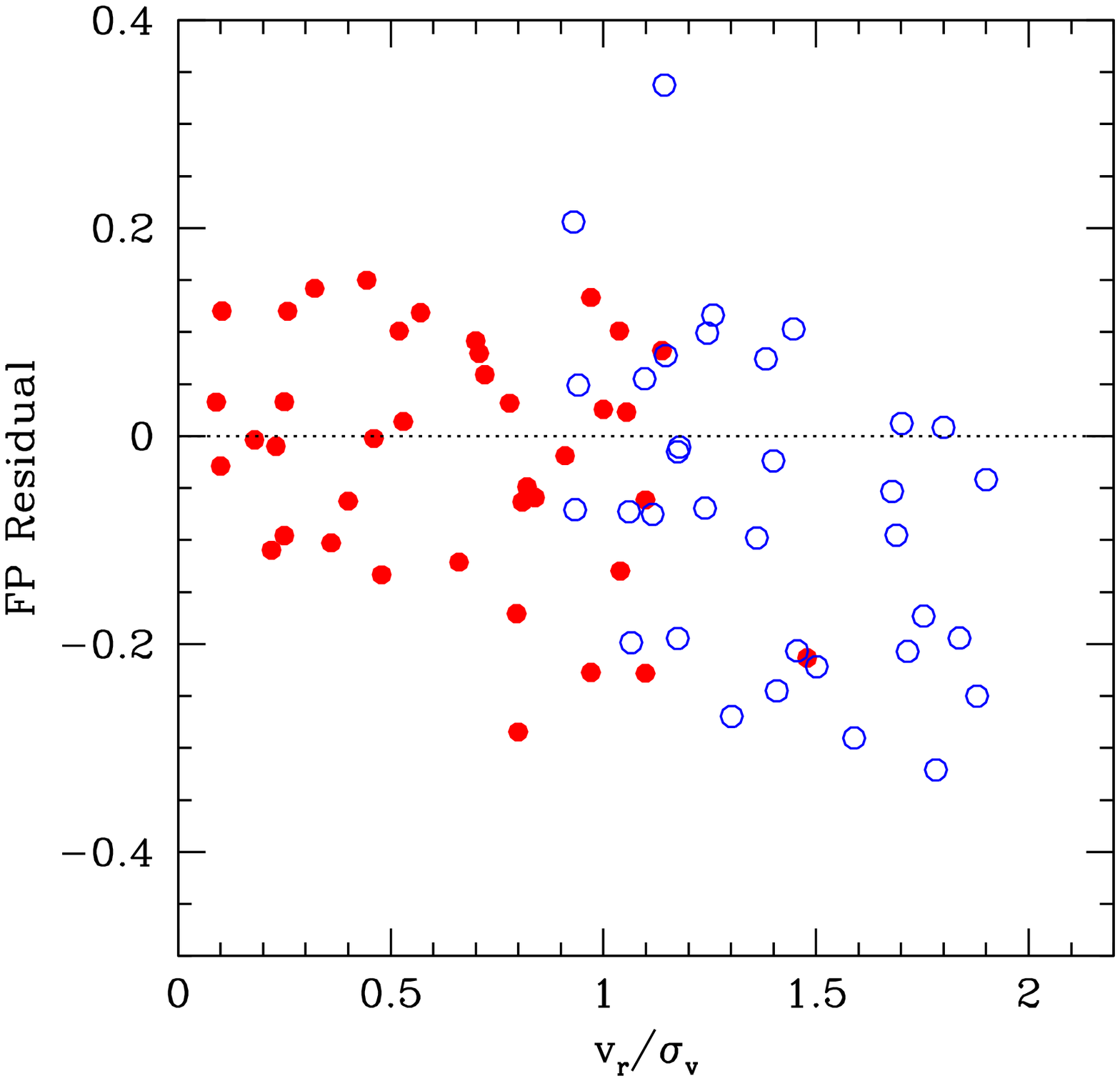}
\caption{Deviation from the FP as a function of $v_r/\sigma_v$ for two galaxy samples, ellipticals from \cite{vdm2} (filled red) and S0's from \cite{bedregal} (open blue). As $v_r/\sigma_v$ increases the deviations become systematic and negative, indicating some dependence on $v_r$ is needed in the scaling relation. The interpretation of this trend is that the FP is not accounting for the dynamical support provided by rotation and that this omission causes larger deviations as the rotational support becomes more important.}
\label{fig:s0}
\end{figure}

Even ignoring the galaxies with low $\sigma_v$ in Fig \ref{fig:m2l}, 
the FP is only valid for galaxies whose stellar distribution is dynamically supported by stellar random motions. We must therefore search
for a description that is more broadly applicable than the FP. In certain early-type galaxies, 
lenticulars in particular, but also lower luminosity spheroidals \citep{davies, bender}, rotation provides an important additional source of dynamical support. This support is evident in the classic diagram comparing the ratio of the rotational velocity, $v_r$, to the velocity dispersion, $v_r/\sigma_v$, vs. ellipticity \citep{davies}, but also becomes evident when plotting the deviation from the FP (using the parameterization $r_h \propto \sigma_v^{1.2\pm 0.07}I_h^{-0.82 \pm 0.02}$ from \cite{cappellari}) vs. $v_r/\sigma_v$ using data for ellipticals \citep{vdm2} and S0's \citep{bedregal} in Figure \ref{fig:s0}. 

As galaxies become more rotationally supported, the use of only $\sigma_v$ to measure their dynamical support leads to a systematic deviation from the FP. Therefore, a natural extension of the FP invokes a combination of $\sigma_v$ and $v_r$, referred to here as $V$, to describe this support. The commonly suggested combination is expressed as $V =  \sqrt{v_r^2/\alpha + \sigma_v^2}$, where $\alpha$ is a parameter that is determined by the internal structure of the system. Unfortunately, 
the existing data are insufficient to discriminate between choices of $\alpha$ suggested previously \citep{burstein,weiner, kassin, zzg}.  A standard choice is $\alpha = 2$, although $\alpha=3$ is also acceptable \citep{weiner} and \cite{zzg} fit the data to argue for $\alpha = 2.68$. The allowed range in values of $\alpha$ is related to the unknown nature of the gravitational potential and the tracer particle distribution function. Within an isothermal potential, if the orbits are isotropic, then $\alpha = 2$, while for other potentials and orbital anisotropies the value of $\alpha \ne 2$. Here, we will simply adopt $\alpha = 2$.
Although the value of $\alpha$ is critical for certain applications, such as determining whether subtle variations in the value of $\Upsilon_h$ exists between ellipticals and S0's, it is not so for the discussion here. In general, for pressure supported systems $V$ reduces to $\sigma_v$, which is what we used in our discussion of giant ellipticals, and for purely rotationally supported galaxies $V$ reduces to $v_r/\sqrt{\alpha}$, which is what we will use for late type galaxies and then only affects the normalization of the late-types relative to the early types on the FM.  We now proceed to discuss the inclusion of late-type galaxies in this context\footnote{In principle, one can determine the value of $\alpha$ by requiring agreement between early and late type galaxies for a single value of the constant in Eq. \ref{eq:walker}. However, in practice because of the dichotomy between FP and TF studies, the samples of early and late type galaxies have been observed and analyzed differently and small differences in photometric systems or analysis technique are sufficient to explain the offsets found.}.

It is evident from Figure \ref{fig:s0} that some of the scatter in the FP, even among ellipticals, comes from neglecting rotational support and that studies that use the FP to search for stellar population differences among early-types, for example those searching for differences among galaxies in different environments, could face systematic errors if the degree of rotational support differs in systematic ways. This issues could even trickle down to the interpretation of studies that do not use the FP, but use $\sigma_v$ as an indication of mass \citep{zhu}.

\subsection{Onward to and Beyond the Tully-Fisher Relationship}

Unlike the FP, the TF does not involve a radial scale and it has
resisted improvement (i.e. a reduction in scatter) via the addition of other structural parameters \citep{aaronson}. The lack of scale dependence is a bit puzzling in that one would expect a galaxy that is physically twice as large as another, but with the same rotation velocity, to have twice as much mass and hence twice the luminosity. As such, the existence of the TF implies that the larger galaxy must have a commensurate lower surface brightness
(so that it does not have a higher total luminosity) --- and that this tradeoff must be nearly balanced to avoid the expected scale dependence. Some hints of a radial dependence are appearing with larger datasets and better data. While these studies suggest that this balance is not exact \citep{courteau, saint},  we again see that the existence of scaling relations  requires that the process of galaxy formation results in systematic
packing of the luminous baryons. For disks, analytic formation models have long advocated a connection between the specific angular momentum of the baryons and surface density or disk size \citep{fall,dalcanton}, which goes part of the way to explaining this balance. Furthermore, systematic behavior between rotation curves and dark matter halos has been identified \citep{salucci07}, which demonstrates the existence of connections between the formation of the entire galaxy and the properties of the inner, observable component.

\begin{figure}
\plotone{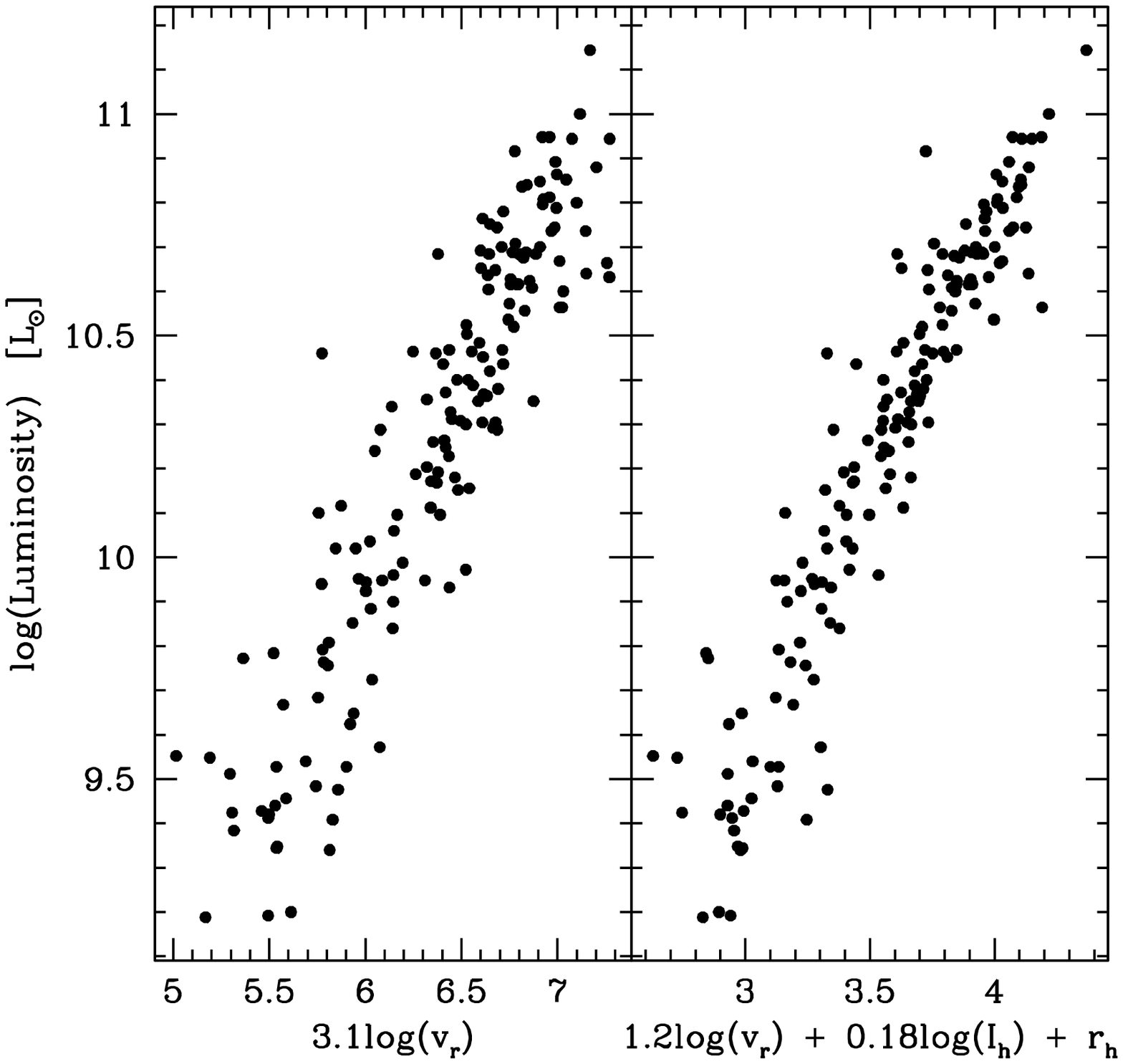}
\caption{Predictions of galaxy luminosity from TF and our re-written FP for disk galaxies. We compare the predicted luminosities (x-axis) to measured luminosities using the galaxies Hubble distance (y-axis) for galaxies from the \cite{pizagno} study using their best-fit TF relationship (left panel) to and that using our re-expression of the \cite{cappellari} FP relation and the use of the $V$ parameter (right panel). The scatter is evidently smaller in the right panel (0.75 in the left one, 0.47 in the right), showing that including a radial scale, as is done in the FM, adds significant information.}
\label{fig:comp}
\end{figure}

Because we seek a universal scaling relation, I avoid the TF for now and assert that Eq. \ref{eq:walker},
inclusive of the combined kinematic support from $v_r$ and $\sigma_v$, should apply to late type galaxies. I therefore rewrite
the FP as an expression for the luminosity, $L$, for comparison to TF, again using the \cite{cappellari} parameterization of the FP,
\begin{equation}
L = 1.2 \log V + 0.18 \log I_h + r_h + Constant
\label{eq:diskfp}
\end{equation}
In Figure \ref{fig:comp},
I compare the predictions of $L$ from the TF \citep[i.e. using the circular velocity and the best fit TF relation presented by ][]{pizagno} and from the rewritten FP (Equation \ref{eq:diskfp}) with the luminosities inferred using distances calculated from their recessional velocities for disk galaxies with $V > 100$ km sec$^{-1}$ (because the FP applies only to such galaxies). 
I use the \cite{pizagno} study because it is one of the few TF samples that includes measurements of $r_h$. The scatter is visibly lower in the re-written FP relation than in the TF relation (0.47 vs. 0.75), demonstrating that the FM parameterization is at least as good as the best-fit TF.  Previous studies that searched for a scale dependence may have had a difficult time finding one because they either used different radii than $r_h$, such as the disk scale length or an isophotal radius, or faced additional noise from not using the rotational velocity at or near $r_h$.  Although for disk galaxies, with their mostly circular orbits, one can argue that $M \propto rv_r^2$ should be a fairly good approximation regardless of which of the standard fiducial radii (disk scale, isophotal, half light) is chosen \citep[see][for empirical support for this claim]{yegorova}, the only one of these radii that has a direct connection to the total luminosity is the half light radius. 

To see how the TF is a subset of the relationships allowed by Eq. \ref{eq:walker}, I 
begin with the TF relation
\begin{equation}
L = Av_r^\delta,
\end{equation}
where $A$ and $\delta$ are constants, rewrite the expression
using the definition of $I_h$, and rearrange it to be in the form of Eq. \ref{eq:fm1},
\begin{equation}
\log r_h = \delta \log v_r - \log I_h + Constant.
\label{eq:tf}
\end{equation}
Again there is a strong similarity to Eq. \ref{eq:walker} and the validity of both equations for giant spiral galaxies
suggests that $V = v_r$ and that $\Upsilon_h \propto v_r^{2-\delta}$ for these spirals. Because empirically $\delta \sim 3 - 4$ \citep{pizagno,courteau}, $\Upsilon_h$ decreases as $v_r$ increases. Therefore, it is wrong to accept that the TF  applies in general because, like the FP, it too faces a fundamental conceptual flaw if extended beyond the class of galaxy from which it was derived. The FP predicts unphysical values of $\Upsilon_h$ for low $\sigma_v$ systems. The TF predicts unphysical values of $\Upsilon_h$ for high $v_r$ galaxies.  Therefore, just as in the case of the FP, some curvature in the relation is necessary for disk galaxies. Suggestions of nonlinearity in the TF relation have existed for over 20 years \citep{mould, persic}. Neither the FP nor TF hold across all galaxy masses, and a more complex relation is needed between $\Upsilon_h$ and $V$.

\subsection{Closure on a Universal Galaxy Kinematic Scaling Relation}

Returning to Equation \ref{eq:fm1}, we see that the three ingredients involved in arriving at a universal galaxy scaling relation are 1) the virial theorem, 2) the robustness of mass estimates at the half light radius, which affirms that the virial coefficients are relatively independent of the detailed structure and kinematics of each galaxy, and 3) the relationship between $\Upsilon_h$ and the directly measurable structure parameters $(r_h, V_h, I_h)$.  The first is a natural expectation for collapsed systems, particularly since $r_h \ll r_{virial}$. The second has been demonstrated by \cite{walker} and \cite{wolf} but has  been evident in the low empirical scatter of the scaling relations for decades. The last remains perhaps the most mysterious because it involves a connection between the efficiency of star formation and the packing of those stars within dark matter halos of galaxies. Understanding this behavior is particularly challenging given that the values of $\Upsilon_h$ for galaxies vary between values of $\sim 1$ and $\sim 1000$. This behavior lies at the core of 
understanding galaxy formation.

The classic scaling relations, FP and TF, implicitly adopt a power-law scaling between $\Upsilon_h$ and the observables. This description fails as we proceed to either low (Figure \ref{fig:m2l}) or high $\sigma_v$ galaxies. The power-law assumption must be abandoned.  A more complicated manifold relates the observables to $\Upsilon_h$ and was named the Fundamental Manifold \citep{zgz} in reference to its antecedent the FP. In the interest of completeness, I present one specific parameterization of that surface from \cite{zzgb}
\begin{eqnarray}
\Upsilon_h &=& 1.49 - 0.32\log V_h - 0.83 \log I_h + 0.24 \log^2 V_h + \nonumber \\
&&  0.12 \log^2 I_h - 0.02 \log V_h \log I_h.
\label{eq:m2l}
\end{eqnarray}
However, because there is yet no set of homogeneous data that includes internal kinematics and spans all galaxy types and luminosities, the derivation of this surface remains preliminary and varies quantitatively depending on what sample is used. Additionally, there is no physical motivation
for why the surface has any particular functional form. As such, larger samples that cover the full parameter space are critically needed. Should subsequent investigations define a completely different functional form for $\Upsilon_h$, that finding would not invalidate any of the discussion presented so far other than Equation \ref{eq:m2l}.

Similar curvature is seen in the behavior of $\Upsilon$ in efforts to match the halo and stellar mass functions \citep{vdb,yang} and lensing mass measurements and luminosities \citep{hoekstra,mandelbaum}. Because the relationship is currently only empirical, the finding that all stellar systems can be modeled with a simple, smooth functional form is perhaps surprising --- but it hints at rules governing galaxy formation that are currently not fully understood. 

\subsection{What Comes Next?}

On the observational front, the refinement of the FM requires a homogeneous dataset that includes all galaxy types and luminosities and which is volume representative (not necessarily complete). Such data would enable 1) a  determination of $\alpha$, 2) tests of whether the scatter increases for certain classes of galaxies, which would address whether the virial coeffciients are sufficiently constant across galaxy types, and 3) a determination of how the surface relating $\Upsilon_h$ to $(V_h, I_h)$ is populated to identify additional constraints on galactic structure. Care should be taken to minimize sources of photometric scatter by choosing passbands that minimize variations arising from stellar population variations and extinction. The sample should consist of galaxies with independent distance estimators so that distance uncertainties contribute minimally to $r_h$. Currently, the distances are calculated using the recessional velocities and an adopted value of the Hubble constant.  2-D kinematics would help address issues related to inclination corrections for rotation speeds and help in the measurement of $\sigma_v$ and $v_r$ in systems where both contribute noticeably to the dynamical support. The sample might also include more systems where $\Upsilon_h$ can be estimated in an independent manner, for example through the use of gravitational lensing.  Finally, and particularly important for questions related to the nature of the luminous baryon distribution, it is necessary to have independent 
estimates the stellar mass-to-light ratio, $\Upsilon_*$. One such approach at measuring $\Upsilon_*$ is to use infrared luminosities and colors \citep{meidt,eskew}, although these estimates depend on stellar population models, that have their own set of uncertainties. Should these uncertainties be sorted out in subsequent work, we would be able to use the FM and these estimates to uncover the dark matter fractions for all galaxies within $r_h$ in a way now done only with more sophisticated modeling on smaller samples of galaxies \citep{cappellari12}.

\section{Discussion}

\subsection{The Bifurcation of Stellar Systems}

The function that describes the relation between $\Upsilon_h$ and ($V_h$,$I_h$) is roughly parabolic along the $V_h$ axis and power-law like along the $I_h$ axis (Equation \ref{eq:m2l}). The general shape places constraints on how baryons settle and become stars within dark matter potentials. However, an important aspect that I have mostly ignored so far is how this surface is populated. In particular, there is a bifurcation in the population at low $V_h$ where two branches develop,  one heading upward to large values of $\Upsilon_h$, populated by the Local Group dwarf spheroidals \citep{zgz,tollerud,salucci12} and ultrafaint galaxies \citep{zzgb,willman}, and another heading downward to low values of $\Upsilon_h$, populated by ultracompact dwarf galaxies and star clusters \citep{mieske,zzgb}. The question is whether this bifurcation points to a problem with the scaling relation \citep{forbes} or whether it is pointing to a fundamental difference between systems with and without dark matter halos \citep{zzgb}.

The bifurcation may lie at the heart of a physical distinction between galaxies and star clusters \citep{willman} and the FM may be a way to help explore that division. In particular, it would be of interest to determine whether there are objects that lie between these two branches, or whether the branches are absolute \citep{misgeld}. Unfortunately, strong selection effects come into play. At the extremely low surface brightnesses of the ultrafaint galaxies, these systems require star counts to detect and are therefore not within reach if they lie outside the Local Group. It may be possible, with the next generation of sky surveys, to detect large numbers of such objects even though we are confined to finding only the local ones \citep{tollerud08}. On the high surface brightness side, these objects become confused with stars \citep{hilker,drinkwater,phillipps} and therefore require unbiased redshift surveys to identify \citep[for recent examples see][]{misgeld11,chiboucas}. There are a handful of Galactic objects that might lie between the branches \citep{zzgb} but these systems are all of low mass, highly susceptible to dynamical effects, and may therefore not satisfy the basic requirement of Eq. \ref{eq:fm1}, which is that the system satisfy the virial theorem.

\subsection{Theoretical Work}

The physics of galaxy formation determines the relationship between $\Upsilon_h$ and the structural parameters of 
each galaxy. As such, we desire a theory that will explain the principle observables of each galaxy individually. Many attempts to reconcile the theory of galaxy formation as currently understood to observed scaling relations (either one for spheroids or disks) exist \citep[for some examples see][and references therein]{dutton}. The key to these models is how one follows the complicated physics of galactic feedback and angular momentum transfer between the baryons and dark matter. The models can reproduce general galaxy properties if one artificially imposes how these effects scale with halo mass. Of course, the next question is why such recaling exists. \cite{dutton} conclude that certain processes, such as the angular momentum evolution of galaxy disks, are not yet sufficiently well understood to be modeled sufficiently accurately to reproduce the scaling relations. Empirical scaling relation provide challenging benchmarks for the models. Extending the range of these scaling relations, for example down to ultrafaint galaxies, is invaluable because it places even stronger constraints on hypothesized physical mechanisms. For example, \cite{anderson} and \cite{mcgaugh} used the baryonic TF relation to argue that there are basic conceptual problems with feedback models, while \cite{dutton12} then used this argument to examine how to obtain plausible models for star formation and feedback. This article is not intended as review of theoretical models, but such modeling is critically dependent on the best possible empirical constraints. Because current simulations lack the ability to treat the physical processes in detail and realistically, they are not reliably predictive and must constantly be compared to the available constraints.

\section{Applications of the FM}

Although understanding galaxy formation remains the principal goal behind refining and understanding the FM, the relationship also has various uses that do not require a theoretical underpinning. For example, as with the FP and TF before it, the FM can be used as a distance estimator because $r_h$ is in physical (distance dependent) units.

\subsection{Using the Scaling Relationships for Distance Measurements}

The advantages that the FM provides over its antecedents come from its universality. For example, the FM is applicable to low luminosity local systems, in which individual stars are resolved but neither the FP or TF apply. This capability is critical because certain distance methods require resolved stellar populations, which are beyond our current technology's reach for most of the galaxies that we know satisfy the FP and TF. As such,  the FM can be used to cross-calibrate distance estimators even when those estimators are not found in the same galaxy. For example, Figure 4 (reproduced from Figure 2 of \cite{zzgb}) illustrates why it is difficult to compare distances obtained from surface brightness fluctuations (SBF) and Cepheids if one requires having distance estimates for the same galaxy from the two methods. Likewise, certain methods tend to work over limited distance ranges and are therefore not suitable for comparison with other.
Using the FM as a fiducial, distances obtained using SN Ia, Cepheids, SBF, the luminosity of the tip of the red giant branch (TRGB), circumnuclear masers, eclipsing binaries, RR Lyrae stars, and the planetary nebulae luminosity functions (PNe) were compared by \cite{zzgb}.

\begin{figure}
\plotone{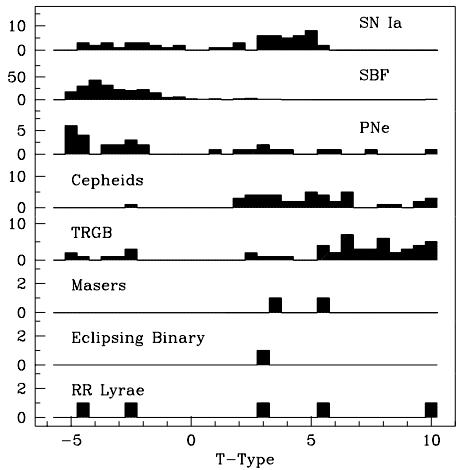}
\caption{Distribution of morphological types as a function of distance measuring technique among a sample drawn from NED-1D. The Y-axis is arbitrarily normalized to 
match scales among subsamples. Differences among the types of galaxies accessible to various methods are clear \citep[adopted directly from Figure 2 of ][]{zzgb}}
\label{fig:cross}
\end{figure}

\begin{figure}
\plotone{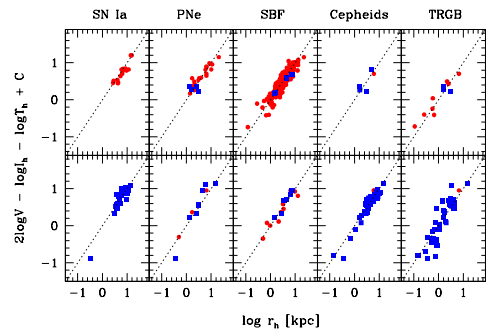}
\vskip -1cm
\caption{FM using measured distances. A comparison of the FM's obtained using different distance estimators. The sample is divided into pressure supported (upper panels) and rotationally supported (lower panels). Furthermore, color and shape codes distinguish galaxy morphologies (blue squares for late type, red circles for early type).
The x axis is log $r_h$ in kpc, and the y-axis is $2 \log V - \log I_h - \log \Upsilon_h - C$, where $V$ is in km sec$^{-1}$, $I_h$ is in $L_\odot/$sq. pc, and $\Upsilon_h$ is in solar units. $C$ is obtained by calibrating to the sample of surface brightness fluctuation distances (SBF) for the pressure supported galaxies. The line is the 1:1 expectation, not a fit to the data  \citep[adopted directly from Figure 7 of][]{zzgb}.}
\label{fig:dist}
\end{figure}

This comparison is done by placing galaxies with independent
distance measures on the FM, using the distance from each particular distance estimator to convert $r_h$ from angular to physical units. 
In Figure \ref{fig:dist} we reproduce Figure 7 from \cite{zzgb} that presents the FM relationship derived using the distance measurements obtained from each type of distance estimator for which there are at least 10 galaxies with all the necessary data. 
The velocity dispersion dominated systems, which have measured velocity dispersions and sometimes also have stellar rotation values, are plotted in the upper panels of the Figure and the
purely rotationally supported galaxies, which have no quoted stellar velocity dispersion and where the inclination-corrected \ion{H}{1} width is adopted as a measure of $v_r$, are plotted in the lower panels.  In general, if there is a stellar velocity dispersion measurement, the galaxy is a pressure supported systems ($v_r/\sigma_v < 1$).  Within each panel a color-code describes morphology, dividing the early and late type galaxies at a T-Type of 1. The zero point of the FM is set using the results from the pressure-supported, SBF sample, which is the largest subsample of galaxies and also predominantly consists of early-type galaxies, which are
less susceptible to extinction and stellar population variations.
Although many more galaxies than those shown have distances measured using at least one of these methods, many of those lack the  measurements necessary to place them on the FM.  The solid lines represent the FM and the only ``free" parameter involved here is the normalization applied to get the mean SBF relation exactly (on average) on the FM.
Normalization or slope errors for any particular distance estimator suggests that there is a problem
with that estimator. Increased scatter suggests a lower precision for that particular distance estimator.

A cursory examination of the panels reveals no serious problems with
any of the distance estimators as applied to any of the galaxy subsamples, even though some perform better than others (either in terms of zero point or scatter). There are a few individual galaxy outliers, although it is often the same galaxy that is an outlier in multiple panels because distances are sometimes available from multiple estimators, suggesting that fault lies not with 
the distance estimate but rather with one of the other parameters that enters the FM. There are no statistically significant differences in zero point between the distance estimators \citep{zzgb}, although the allowed differences are still above the level of precision ($\sim$ percent) that is the goal of current studies. Increasing the sample significantly, which is well within what can be done with a reasonable investment of resources,  will provide stricter limits on potential systematic differences among estimators. 

\cite{zzgb} also examined the FM residuals within a given distance estimator vs. other potential sources of systematic error. For example, they examined the relationship between FM residual and host galaxy metallicity for distances derived using SNe Ia, they found a correlation, confirming previous claims of metallicity-dependent correction to the Ia distances \citep{gallagher, howell, romaniello}.

\begin{figure}
\plotone{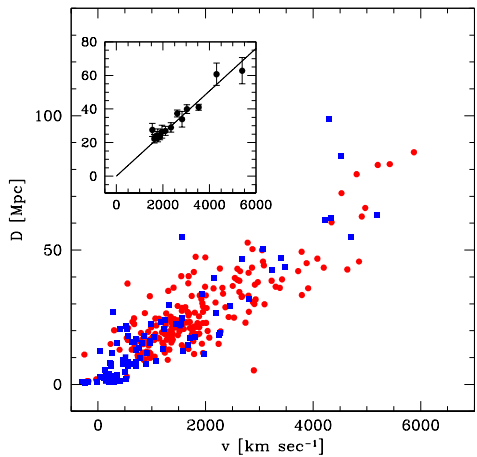}
\caption{Hubble diagram using distances derived using the FM. Pressure supported systems are plotted as red circles and rotationally supported systems as blue squares. Inset shows data at $v > 1500$ km sec$^{-1}$ binned so that each bin contains 10 galaxies. The fitted line is constrained to go through the origin and corresponds to $H_0 = 78 \pm 2$ (random) $\pm 4$ (systematic) km sec$^{-1}$ Mpc$^{-1}$ \citep[adopted directly from Figure 12 of][]{zzgb}}
\label{fig:h0}
\end{figure}

Once established as a distance estimator itself, the FM can be used
for a large set of galaxies. In Figure \ref{fig:h0}, I reproduce Figure 12 of \cite{zzgb} that shows the relationship between recessional velocity and distance obtained from the FM for the same set of galaxies shown in Figure \ref{fig:dist}. Comparing between different normalization of the distances, using the various distance estimators available to them, \cite{zzgb} cite a systematic uncertainty of 4 km sec$^{-1}$ Mpc$^{-1}$ in $H_0$.
The inset shows the result of fitting the relationship for galaxies with $v > 1500$ km sec$^{-1}$, $3\sigma$ outliers excluded, binned by 10, with the fit forced  through the origin.  The low $v$ region is excluded to minimize the effect of local flows. They find a best fit slope corresponding to $H_0 = 78 \pm 2$ (random) $\pm$ 4 (systematic) km sec$^{-1}$ Mpc$^{-1}$. The estimate of the systematic uncertainty does not include a variety of potential problems that they ignored  (modeling of bulk flows, internal extinction corrections, adjustment for potential biases in the galaxy sample, etc.), but this calculation was done primarily as a plausibility exercise to demonstrate the use of the FM.

One of those sources of uncertainty, internal extinction, would be mitigated by going to the infrared, as shown most recently by \cite{freedman11} for {\sl Spitzer} wavelengths where the scatter about the TF relation is
reduced from 0.43 in the extinction-corrected $B-$band data to 0.31 in the non-extinction-corrected
3.6$\mu$m data. One avenue for advancement is therefore the use of large IR photometric galaxy samples \citep{sheth}.

\begin{figure}
\plotone{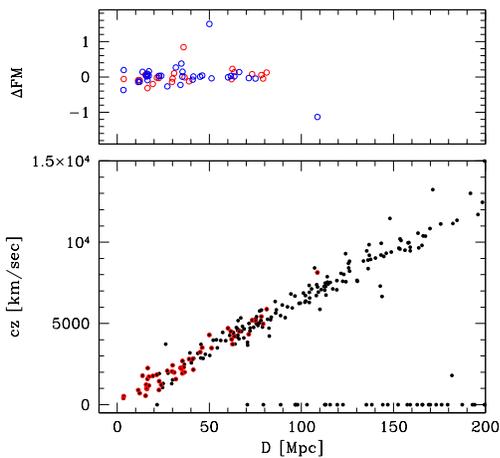}
\caption{SN Ia distances and the FM. The upper panel shows the residuals about the FM relation ($\Delta$FM) for galaxies with SN Ia distances. Blue and red symbols denote rotationally and pressure supported systems respectively. The bottom panel shows the Hubble diagram for galaxies with SN Ia distances available in the NED 1-D database. The galaxies with all the necessary data to be placed on the FM are highlighted in red. There are numerous galaxies with existing SN Ia distances that could be added to the FM with reasonable observational effort. The most noticeable outliers from the FM are also outliers in the Hubble diagram.}
\label{fig:todo}
\end{figure}

The most straightforward way for advancement is to obtain kinematic measurements for a large number of galaxies with existing distance measurements. In Figure \ref{fig:todo} I present a subset of galaxies with SNe Ia distance measurements and highlight those for which all
the necessary data are available. An interesting indication of how these data may help refine distances is provided by two outliers from the FM (at 50 Mpc and the one datum beyond 100 Mpc), which are both
outliers in the SNe Hubble diagram. That these galaxies are outliers in both panels suggests that there is a problem with the distance. With larger samples, one could then attempt to find the cause for the distance anomaly by comparing with other characteristics, either those of the SNe itself or those of the host galaxy. Refining the SNe distances has broad implications for cosmology.

\subsection{Using the Scaling Relationships for Mass Measurements}

The stellar mass of a galaxy is intricately connected to a variety of galaxy properties: environment \citep{kauffmann04}, metal abundance \citep{tremonti}, star formation history \citep{bundy}, and  dark matter halo mass \citep{vdb},  to name a few. However, the stellar mass estimates we rely on are quite crude and potentially rife with systematic errors. 
A seemingly straightforward way to estimate the stellar mass is to use measurements of the stellar populations, such as color, to estimate $\Upsilon_*$  and to convert the luminosity into a mass. However
those estimates depend on three factors that are not well understood: 1) the galaxy's star formation history, which can be quite complex \citep[see for examples][]{harris04,harris,eskew11,weisz} 2) stars' behavior during the phase(s) of their life at which they are at their most luminous, which is problematic \citep{maraston,conroy,mcquinn,melbourne} and 3) 
the initial distribution of stellar masses (the initial mass function or IMF), which is a well-known unresolved problem \citep{bastian}. An alternative method is to dynamically measure the mass and, if necessary, make a correction for the amount of dark matter present.  The FM enables us to measure the
total masses within $r_h$. 

An interesting set of stellar systems to examine are stellar clusters, which presumably contain no dark matter and are single age populations. As such, the estimates of $\Upsilon_h$ obtained using the FM provide a measurement of the mass-to-light ratio of a stellar population, $\Upsilon_*$, of a certain age. 
A large compilation of  Local Group stellar cluster data was published by \cite{mclaughlin}, although the velocity dispersions available at the time of that study were either for the older ($>$ 10 Gyr) clusters or the very  young ones ($<$ 100 million yr). \cite{z12} filled in this range by observing clusters over the full range of ages. 

Discrepancies between models and observations of galaxies, when found, are often attributed to deviations in the IMF from the adopted prescription \citep[for some recent examples see][]{vandokkum08,dabringhausen09,treu10,vandokkum,dutton12a} rather than problems with either the star formation history or stellar evolutionary models. However, direct measurements of the initial mass function are difficult for various reasons \citep[see][for a review]{bastian}, particularly over the full range of environments and conditions. 
These clusters provide one of the most direct tests of the IMF.

\begin{figure}
\epsscale{0.7}
\plotone{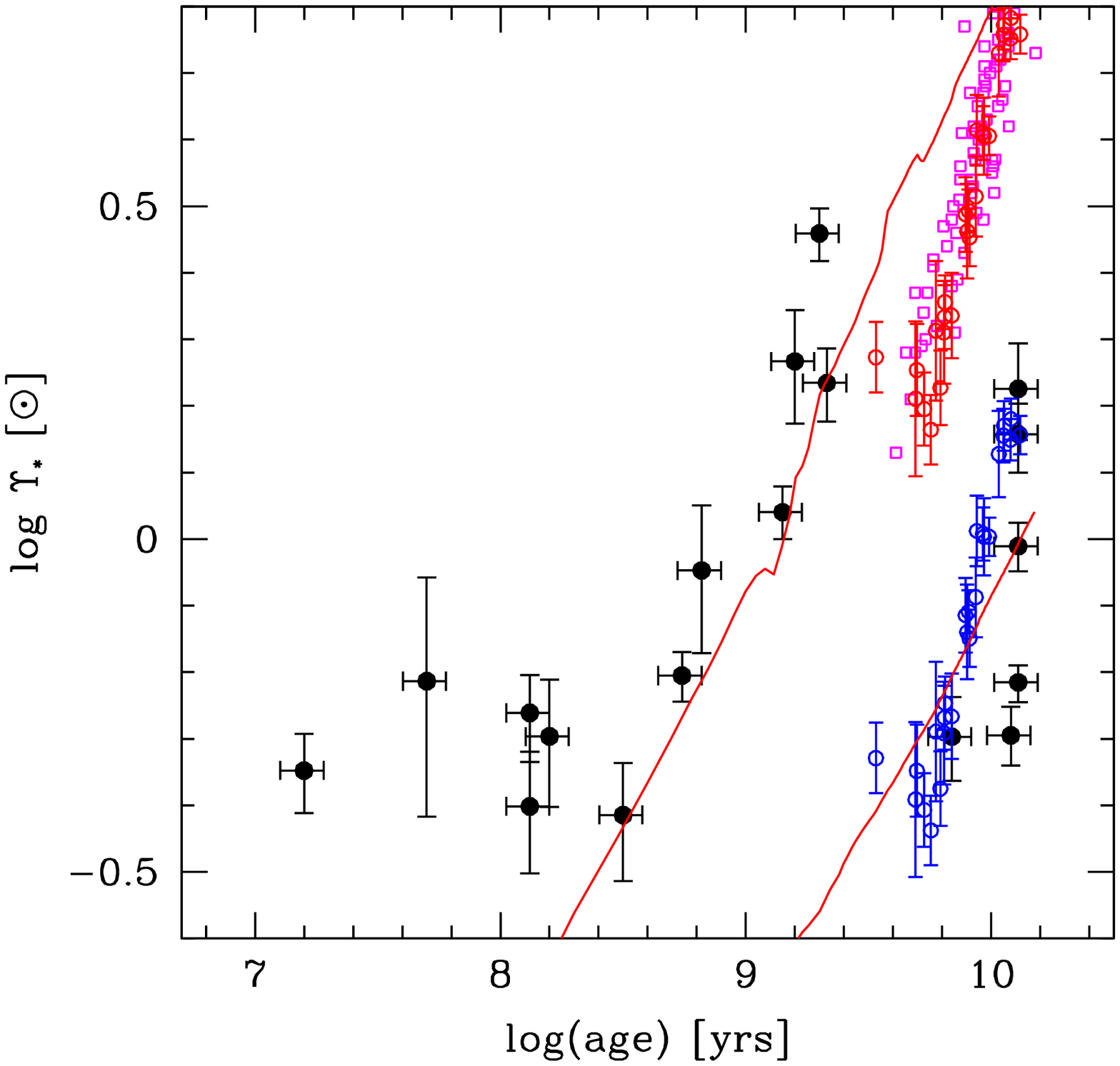}
\caption{The evolution of $\Upsilon_*$ as inferred from early type galaxies and comparison to models and our cluster data. The stellar cluster data are plotted in solid circles (black), the results obtained using the results from \cite{vd07} are plotted with open circles (red) when normalized to match the results from the sample of \cite{graves}, which are plotted with squares (pink).  Also shown are the \cite{vd07} data assuming a dark matter fraction of 80\%, so as to qualitatively match to the stellar cluster data, in open circles (blue). The dark matter fraction is likely to be somewhere between 0 and 80\%. The upper solid line represents the values of $\Upsilon_h$ from a PEGASE model of a population with an instantaneous burst at $t=0$ and a Salpeter IMF, while the lower represents a model with a light-weighted Kroupa IMF (adopted directly from Figure 9 of \cite{z12}).}
\label{fig:m2lcl}
\end{figure}

The principal empirical result presented by \cite{z12}, the relationship between $\Upsilon_*$ and age, is reproduced in Figure \ref{fig:m2lcl} (their Figure 9 reproduced here). The naive expectation, that $\Upsilon_*$ rises continually with age is evidently not reproduced by the data. Internal dynamics, principally two-body relaxation which causes preferential loss of low mass stars over time \citep{spitzer, kruijssen08}, plays a role here, but it is insufficient to explain the large drop in $\Upsilon_*$ between a few and 10 Gyr. \cite{z12} support this claim by applying the dynamical models of \cite{anders}.  A second possible reason for why the naive expectation may not be met is the influence of binary stars, which can artificially inflate the observed velocity dispersion of a system \citep[for examples see][]{minor,gieles10b}. Again the effects are considered and found to be too weak to explain the observations or too contrived, requiring unknown an binary population that affects the kinematics of $\sim~ $1 Gyr populations but not those of age $>$ few Gyr. Variations in the initial mass function among these systems is
a viable explanation. 

In the same Figure, \cite{z12} also included two model tracks for the evolution of simple stellar populations. First, the plotted values of $\Upsilon_*$ for the stellar clusters are corrected for the effect of internal dynamical evolution using the results of the \cite{anders} models.
They plot the results of PEGASE models \citep{pegase} using a \cite{salpeter} IMF, spanning from 0.1 to 120 M$_\odot$, with default stellar mass loss and binarity parameters, and metallicity matching the mean of the young clusters ($-0.4$). These model values are not renormalized in any way and yet do an acceptable job of reproducing the trend seen in the younger clusters, for clusters of 8 $< $log(age [yrs]) $<$ 9.4. Discrepancies at younger ages can be ignored due to the likelihood that these clusters are not relaxed \citep{goodwin}. The second model is that of a lightweighted \cite{kroupa01} IMF, where $\sim$ 50\% of the mass is removed (lightweighted) to produce the match in the Figure. A \cite{chabrier} IMF works similarly. One interpretation of this Figure 
hypothesizes the presence of two IMFs for star clusters, 
one IMF being primarily, but not exclusively, appropriate for older, metal poor clusters and the other for primarily, but not exclusively, for younger, metal rich clusters. The young (log(age [yrs])$<$9.5) clusters are well-described by a bottom-heavy IMF, such as a Salpeter IMF, while the older clusters are better described by a top-heavy IMF, such as a light-weighted Kroupa IMF, although neither of these specific forms is a unique solution. Ongoing work will at least double the sample size, but  we will eventually need to obtain high resolution deep imaging of the intermediate age clusters to confirm that these have a bottom heavy mass function. While some work has been done with {\sl HST} \citep{glatt}, deeper, less ambiguous results are necessary to definitively prove or disprove this interpretation. Such work is within reach of adaptive optics on large telescopes.

In addition to stellar clusters,  the FM can also be applied to estimate the masses (within $r_h$) of galaxies across
cosmic time. In particular, this approach can test the hypothesis that large deviations from stellar evolutionary models, rather than IMF variations, are responsible for sudden drop in $\Upsilon_*$ after a few  Gyr. Using measurements of the ages and structural parameters of two independent sets of early type galaxies, \cite{z12} compared the derived values of $\Upsilon_h$ to those derived for the clusters. Such comparisons constrain both $\Upsilon_*$ and the fraction of the mass in the form of dark matter within $r_h$. They used both a study of Sloan Digital Sky Survey (SDSS) local galaxies presented by \cite{graves} and a
study of galaxy cluster FP measurements presented by \cite{vd07}. The latter study provides only differences in $\Upsilon_h$ as a function of redshift, so global shifts of $\Upsilon_h$ are allowed. The results of those studies are include in Figure \ref{fig:m2lcl} for comparison to the cluster values (after the \cite{vd07} data were normalized to match the \cite{graves} data). 
An important distinction for the galaxy results, is that  these systems do contain dark matter and that the exact proportion of dark matter within $r_h$ is unknown and likely to vary as a function of the structural parameters \citep[for some examples from the long history of this topic see][]{babul,graham,marinoni,vdb,cappellari,zgz,wolf}.

The comparisons between clusters, galaxies and models provide several interesting results.
The early-type galaxies lie between the extrapolation of the young cluster trend and that of the old clusters. 
Because of the dark matter content of galaxies, the actual value of $\Upsilon_*$ for the galaxies is likely to be lower
than that plotted. This suggests that they will not lie on the track defined by the young clusters, which is already slightly
above the galaxy data\footnote{Careful modeling is necessary to reach robust conclusions because the galaxies are not necessarily single age populations. If an old galaxy has a small population of younger stars, those stars will lower the effective $\Upsilon$ of the galaxy.}. On the other hand, unless the dark matter fraction is quite large ($\sim 80$\%, the blue points in the Figure), the
ellipticals will not resemble the old clusters either. Although the possibility of such a large DM fraction is not excluded, we may be seeing 
the effects of having a mixture of populations of stars, some with the IMF of the old clusters and some with that of the young clusters. However, there is no strong deviation from the general behavior predicted by simple stellar evolutionary models and so that does not appear to be the cause of the behavior of $\Upsilon_*$ in the stellar cluster sample.

\section{Summary}

Order implies rules. Rules governing the structure of dynamical systems are the manifestation of underlying physics. Understanding this physics is the overarching goal of the study of galaxy formation and evolution. Is there order among galaxies?

In the early days of modern astronomy, the morphological appearance of galaxies was the principal characteristic that dominated attempts at describing order among galaxies \citep{hubble,sandage}. This approach has been mostly upended by 1) the realization that accretion and mergers drive galaxy evolution and alter morphologies, 2) the dynamical importance of dark matter, which clearly is not a part of a morphological scheme, 3) the increasingly quantitative nature of astronomy and large digital surveys, and 4) the discovery of scaling relations such as the Fundamental Plane (FP)  and Tully-Fisher (TF).  In this review, I have described the development of a scaling relation that is an extended version of the FP, called the Fundamental Manifold (FM), that works on all classes of stellar systems and exhibits a scatter comparable to the more restrictive FP and TF relations. 

The FM relationship depends on three distinct conditions being satisfied: 1) that the virial theorem is applicable, 2) that the derived mass enclosed within the half-light radius, $r_h$, be at most only weakly dependent on the distribution function of the tracer particles and the gravitational potential, and 3) that the mass-to-light ratio within $r_h$, $\Upsilon_h$, depend on at most the observed parameters $V_h$ and $I_h$. The first is as expected because we are working with systems that are virialized ($r_h \ll  r_{200}$). The second is somewhat surprising, but has now been verified at least for spheroidal systems, which are the ones that are most likely to have large variations among distribution functions, by the work of \cite{walker} and \cite{wolf}. The last still requires explanation, although it is manifestly a critical component of a complete theory of galaxy formation and evolution.

Going forward there are several areas that will yield immediate returns. First, homogenous samples across galaxy type with detailed kinematics so that there is uniform, high quality data across the parameter space are critical to testing Equation \ref{eq:m2l}, measuring the parameter that relates velocity dispersion and rotation, $\alpha$, and determining the intrinsic scatter about the relation as a function of galaxy mass and morphology. Second, an improved understanding of the initial mass function is key to accurate and precise calculations of $\Upsilon_*$. Determining $\Upsilon_*$ is necessary to uncovering the behavior of dark matter as a function of mass and galaxy type. Third, direct application of the FM to a number of topics, including distance determinations, will yield results even if a deeper understanding of galaxy formation remains elusive in the short term. The FM, like the FP and TF before it, should help propel research forward in a wide range of topic areas.

\begin{acknowledgments}
The author acknowledges the close collaboration with Ann Zabludoff and Anthony Gonzalez during which most of the material reviewed here was developed.
The author acknowledges financial support for this work from a
NASA LTSA award NNG05GE82G and NSF grant AST-0307482 as well as the hospitality of NYU CCPP during his many visits.
\end{acknowledgments}

{}

\end{document}